\documentclass[apl,aps,10pt,twocolumn,floatfix]{revtex4-1}
\usepackage{amsmath}
\usepackage{amssymb}
\usepackage{bm}
\usepackage{graphicx}
\usepackage{upgreek}
\usepackage{longtable}
\usepackage{color}
\makeatletter
\DeclareRobustCommand*\uell{\mathpalette\@uell\relax}
\newcommand*\@uell[2]{
  \setbox0=\hbox{$#1\ell$}
  \setbox1=\hbox{\rotatebox{10}{$#1\ell$}}
  \dimen0=\wd0 \advance\dimen0 by -\wd1 \divide\dimen0 by 2
  \mathord{\lower 0.1ex \hbox{\kern\dimen0\unhbox1\kern\dimen0}}
}

\begin{document}
\title{First ionization potential measurements using laser-induced breakdown spectroscopy}
\author{Ahsraf M. EL Sherbini} 
\affiliation{Laboratory of Laser and New Materials, Faculty of Science, Cairo University, Giza, Egypt}
\author{Mohamed M. EL Faham}
\affiliation{Department of Engineering Basic Science, Faculty of Engineering, University of Benha, Banha, Egypt   }
\author{Christian G. Parigger}
\email{Corresponding author: cparigge@tennessee.edu}
\affiliation{University of Tennessee Space Institute, Center for Laser Applications, \\
411 B.H. Goethert Parkway, Tullahoma, TN 37388, USA}

\date{\today}

\begin{abstract}
The first ionization potential of neutral atoms is determined from thresholds of laser-induced optical breakdown. Bulk material ablation plasma of aluminum, silver, lead, indium and copper is created in laboratory air
with focused, 5-ns pulsed Nd:YAG, 1064 nm IR radiation. At fixed spot size of 2 $\pm$ 0.1 mm, the laser fluence is varied from
16 to 3 J/cm$^2$. The first ionization potentials of the lines Al I 396.2, Ag I 520.9, Pb I 405.8
and 406.2, In I 410.2 and Cu I 515.3 nm are measured to amount to 5.9 $\pm$ 0.2, 7.6 $\pm$ 0.3, 7.4 $\pm$ 0.2, 5.8 $\pm$ 0.1 and 7.7 $\pm$ 0.2 eV, respectively.
\end{abstract}
\maketitle

\section{Introduction}

The generation of laser-induced plasma requires sufficient irradiance to initiate the avalanche type optical breakdown process \cite{Ref1,Ref2,Ref3,Ref4,Ref5,Ref6,Ref7,Ref8,Ref9}. Plasma is described as the fourth state of matter and is mainly composed of four different species: Atoms, electrons, ions and radiation field that are distributed according to well-known equilibrium distribution functions. The minimum energy flux required for plasma formation is called threshold fluence, $\upvarphi_{\rm th}$ (J/cm${}^{2}$). The fluence thresholds \cite{Ref5,Ref6,Ref7,Ref8,Ref9} depend on thermal parameters of the material that include density, latent heat of vaporization, coefficient of thermal conductivity and specific heat as well as on laser excitation wavelength and ionization energy \cite{Ref10,Ref11,Ref12}. In this work, bulk metallic targets are investigated to infer the ionization potentials \cite{Ref13} of selected lines of aluminum, silver, lead, indium and copper.


\section{Laser-induced threshold}
At the laser irradiance threshold, the amount of laser energy just needed to vaporize the target material in the laser focal volume is given by the thermal term \cite{Ref5}, $\upvarphi _{\rm th}^{\rm thermal}$,
which is constant for each material and can be calculated from classical material-dependent properties \cite{Ref1,Ref2}. However, in order to ionize this vapor, an additional laser-dependent term \cite{Ref1,Ref2},
\begin{equation}
\upvarphi_{\rm th}^{\rm laser} = {\rm C}\ \frac{\upvarepsilon_{\rm i} }{\uplambda_{\rm laser}^{2} }  {\rm \uell}_{\rm T},
\end{equation}
needs to be considered. Here, ${\rm \uell}_{\rm T} $ is the thermal conduction length (m), $\uplambda_{\rm laser}^{} $ is the laser wavelength  and $\upvarepsilon_{\rm i} $ is the first potential ionization energy. The threshold term for the laser fluence, $\upvarphi_{\rm th}^{\rm laser}$, originates from the time-averaged ponderomotive energy for linear polarization \cite{Mulser,Lotz}. The proportional relation of the threshold laser fluence with the inverse square laser wavelength and linear thermal conduction length was already studied for nano-materials of different sizes \cite{Ref1,Ref2,Ref3,Ref4}.
The constant C is composed of a combination of electromagnetic constants \cite{Ref1}, including ${\rm m}_{\rm e}$, $\upvarepsilon_{\rm o}$, $\rm c$ and $\rm e$ denoting mass of the electron, vacuum permittivity, speed of light and elementary charge, respectively,
\begin{equation}
{\rm C} = 8\uppi ^{2} {\rm m}_{\rm e} \upvarepsilon_{\rm o} {\rm c}^{2} /  {\rm e}^{2} = 2.235 \times 10^{15} \left({\rm m}^{-1} \right).
\end{equation}

Recent work
discussed measurements of thresholds, $\upvarphi_{\rm th}$, and confirmed that the complimentary laser-dependent contribution, $\upvarphi_{\rm th}^{\rm laser}$, needs to be be added to describe fluence thresholds of laser-induced plasma at the surface of bulk material \cite{Ref1,Ref2},
\begin{equation} \label{GrindEQ__1_}
\upvarphi_{\rm th}= \upvarphi _{\rm th}^{\rm thermal} +2.235 \times 10^{15} \frac{\upvarepsilon _{i} }{\uplambda_{\rm laser}^{2}} {\rm \uell}_{\rm T}.
\end{equation}
The thermal contribution, $\upvarphi _{\rm th}^{\rm thermal}$, amounts to
\begin{equation}
\upvarphi _{\rm th}^{\rm thermal} =\uprho {\rm L}_{\rm v} {\rm \uell}_{\rm T},
\label{equation1}
\end{equation}
and the thermal conduction length \cite{Ref1,Ref2,Ref5}, ${\rm \uell}_{\rm T} $, can be expressed as
\begin{equation}
{\rm \uell}_{\rm T} =\sqrt{\upkappa_{\rm T} \uptau_{\rm laser} / \uprho {\rm C}_{\rm p}} \ \ ,
\end{equation}
\noindent where $\upkappa_{\rm T} $ is the thermal conductivity coefficient of the material (W/m K), $\uprho $ is the density (kg/m${}^{3}$), $\uptau_{\rm laser} $ is the laser pulse duration (s), ${\rm C}_{\rm p}$ is the specific heat at constant pressure (J/kg K) and ${\rm L}_{\rm v}$ is the latent heat of vaporization in (J/kg).

For the evaluation of the ionization energies, Eq. \eqref{GrindEQ__1_} is suggested and applied in this work, and it can be expressed in frequently encountered units for laser fluence, wavelength and thermal conduction length,
\begin{equation} \label{GrindEQ__2_}
\upvarepsilon_{\rm i} \left({\rm eV}\right)=27.9 \left[\upvarphi_{\rm th} -\upvarphi_{\rm th}^{\rm thermal} \right]({\rm J/cm}^2)\frac{\uplambda_{\rm laser}^{2} }{{\rm \uell}_{\rm T} }\left({\rm \upmu m}\right).
\end{equation}
In this formula, $\upvarphi_{\rm th} $ should be larger than the classical thermal vaporization term, $\upvarphi _{\rm th}^{\rm thermal}$. Therefore, at a fixed laser irradiation wavelength and with knowledge of classical properties, one can measure the ionization potential, $\upvarepsilon_{\rm i}$, by decreasing the laser energy to find the fluence threshold, $\upvarphi_{\rm th}$, for plasma generation.

\section{Experimental details}

The experimental configuration for the ionization energy measurements includes a Q-switched Nd:YAG laser device, focusing and attenuation optics, and optical fiber coupled to a spectrometer equipped with an intensified detector \cite{Ref1,Ref2}. The Nd:YAG pulsed laser is operated at the fundamental wavelength, $\uplambda_{\rm laser}$ = 1064 nm, with pulse duration, $\uptau_{\rm laser}$ = 5 ns, delivering an output energy of 470 mJ per pulse.  The laser radiation is focused with a convex lens of 1 m focal length. The focal spots show radii of 2 $\pm$ 0.1 mm, measured using thermal heat sensitive paper (Kentek or Quantel$^\circledR$ heat sensitive paper). The laser fluences are varied in the range of 16 to 3 J/cm$^2$ with a set of calibrated glass attenuators. The variation of the laser radiation is monitored using a 4 \% reflective beam splitter with an absolutely calibrated power meter (Ophir model 1z02165). The emitted radiation from the plasma is collected with an optical fiber of 25 $\upmu$m diameter, connected to the detection system comprised of a spectrograph (SE 200 Echelle Spectrograph) and an intensified charge coupled device (ICCD Andor-iStar DH734-18F).  The camera is used to record the time-resolved spectra in the range of 200 to 1000 nm. The tip of the fiber is  positioned at distance of 18 $\pm$ 2 mm from the plasma expansion axis. Absolute calibration of the detector system is accomplished with a deuterium-halogen light source (Ocean Optics$^\circledR$ DH-2000-CAL) \cite{Ref14}.

In order to obtain clear optical signals from different plasmas, the light emissions were collected over three different laser shots. The strongest spectral line intensity for each material was selected, \textit{e.g.}, the In I at 410.17 nm and Al I at 396.15 nm lines. The isolated lines are selected to record the plasma emission at different irradiation levels, and the signal to background ratio, S/B, is monitored.

\section{Results and discussion}
The target materials used in our work are selected to cover a relatively wide range of expected threshold values from 0.4 for Pb to 3.5 J/cm${}^{2}$ for Cu and diverse thermal properties.

Table \ref{table1} summarizes the thermal and physical parameters.
This table shows that there is an apparent increase of isobaric specific heat of materials except for aluminum which has relatively large heat capacity. There is an apparent increase in the coefficients of the thermal conductivity ($\upkappa_{\rm T}$) and latent heat of vaporization (L${}_{v}$) except for copper and aluminum, respectively. Neither one of the physical quantities is responsible of the threshold of plasma ignition alone but only a combination of these quantities may be responsible for the generation of laser-induced plasma.

\begin{table}[h]
\caption{Thermal and physical parameters of the elements used in this work in SI units for density, $\uprho$, latent heat of vaporization, L$_{\rm v}$, thermal conductivity coefficient, $\upkappa_{\rm T}$, specific heat (isobaric), C$_{\rm p}$ (J/kg K) and recommended wavelength, $\uplambda$ (nm). For indium and aluminum, lines emerging from resonance transitions are indicated. The listed data are taken from the NIST data base \cite{Ref15}.}
\begin{tabular}{p{0.5in}ccccccc} \hline\hline
 & $\uprho$    & L$_{\rm v}$ & $\upkappa_{\rm T}$  & C$_{\rm p}$ & $\uplambda$ \\
& (kg/m$^3$) &  (10$^6$J/kg) & (W/m K)             &  (J/kg K)   &  (nm)        \\ \hline
Pb & 11350 & 1.8  & 35   & 130 & 406.00 \\
In & 7300  & 1.9  & 83.7 & 233 & 410.17 \\
Al & 2700  & 10.8 & 237  & 900 & 396.15 \\
Ag & 10500 & 2.4  & 429  & 237 & 520.90 \\
Cu & 8960  & 4.8  & 401  & 385 & 515.32 \\ \hline\hline
\end{tabular}
\label{table1}
\end{table}
Regarding the lead Pb I line, the midpoint 406.00 nm of the two prominent lines at wavelengths of 405.78 and 406.21 nm is considered because they appear actually as a single line even at lower laser irradiance levels. Stark broadening and choice of resolution instrumental bandwidth (0.12 nm) resulted in overlap of the two lines separated by 0.5 nm.

Table \ref{table2} shows the variation of the combination of thermal conductivity, specific heat, density and laser pulse duration, labeled thermal conduction length, ${\rm \uell}_{\rm T} $, and the heat enthalpy per unit volume, $\uprho {\rm L}_{\rm v} ({\rm J}/{\rm m}^3 )$. The table also shows the thermal term, $\upvarphi _{\rm th}^{\rm thermal}$, see Equation \eqref{equation1}.  The investigated elements are arranged in Table \ref{table1} in ascending order of the predicted threshold laser fluences according to Eq. \eqref{GrindEQ__2_} at the laser wavelength of 1064 nm.

\begin{table}[h]
\caption{Parameters of the investigated elements: Density $\times$ heat enthalpy product, $\uprho {\rm L}_{\rm v}$, thermal conduction length, ${\rm \uell} _{\rm T}$, thermal term, $\upvarphi _{\rm th}^{\rm thermal}$, and predicted thresholds from Equation (\ref{GrindEQ__2_}), $\upvarphi _{\rm th}$.}                                     \begin{tabular}{p{0.5in}cccc} \hline\hline
 & $\uprho {\rm L}_{\rm v}$ & ${\rm \uell}_{\rm T}$ & $\upvarphi _{\rm th}^{\rm thermal}$ & $\upvarphi _{\rm th}$  \\
& (kJ/cm$^3$)        &     (nm)           &  (J/cm$^2$)           &   (J/cm$^2$)   \\ \hline
Pb &   9.7 & 350  & 0.33 & 0.42 \\
In &  14   & 500  & 0.70 & 0.80 \\
Al &  28   & 700  & 1.97 & 2.10 \\
Ag &  25   & 1000 & 2.3  & 2.55 \\
Cu &  43   & 760  & 3.27 & 3.46 \\ \hline\hline
\end{tabular}
\label{table2}
\end{table}

As can be noted from the data
in Table \ref{table2}, there is an increase from Pb to Cu in the thermal conduction length and the density $\times$ heat enthalpy product of the materials except for silver because of the its relatively large thermal conductivity. The thermal contribution, $\upvarphi _{\rm th}^{\rm thermal} =\uprho {\rm L}_{\rm v} {\rm \uell}_{\rm T}$, shows an increase but is
consistently
smaller than the theoretically predicted contribution from the laser fluence, $\upvarphi _{\rm th}^{\rm laser}$, indicated in Eq. \eqref{GrindEQ__1_}.

\begin{figure}[h]
\begin{center}
\includegraphics[width=2.75in]{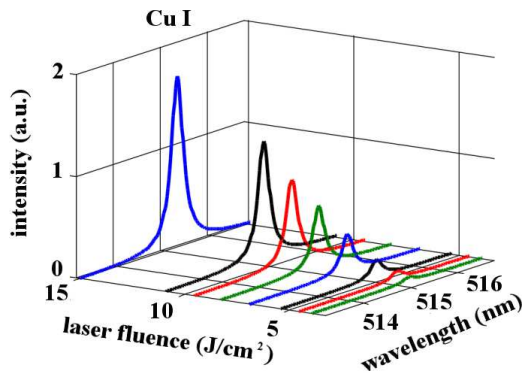}
\end{center}
\caption{Measured spectra of the 515.32 nm Cu I line versus wavelength and laser fluence.}
\label{figure1}
\end{figure}

\begin{figure}[h]
\begin{center}
\includegraphics[width=2.75in]{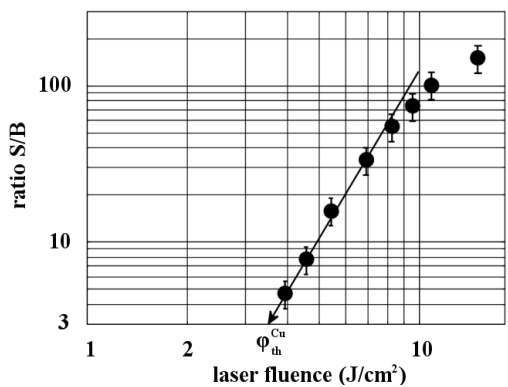}
\end{center}
\caption{Signal to background ratio, S/B, for the 515.32 nm Cu I line and threshold, $\upvarphi^{\rm Cu}_{\rm th}$ = 3.4 J/cm$^2$, at S/B = 3.}
\label{figure2}
\end{figure}

Figure \ref{figure1} illustrates typical recorded copper emission spectra. The integrated spectral intensity is obtained by evaluating the area of the lines after subtracting the continuum or background.
The plasma-threshold fluence  is determined via backward extrapolation at the recommended signal-to-background, S/B, value of three as indicated in Fig. \ref{figure2}.
The log-log plot in Fig. \ref{figure2} shows a  linear decrease in the S/B ratio for smaller  laser fluences in the range of 10 to 3 J/cm$^{2}$.
For larger signal levels, the recorded spectral intensities of the lines tend to saturate. This saturation may be attributed to the self-absorption effects at the relatively large laser energy.

Table \ref{table3} shows the experimentally measured laser threshold fluences corresponding to each element, the measured first ionization energy utilizing the plasma threshold laser fluence, the tabulated standard values for the ionization energies and for the work functions. The work function is defined as the minimum energy required generating free electrons from the surface of a solid.
The work functions of the different elements show nearly constant values in the range of 4.1 to 4.7 eV. However, the discussed model
for laser-induced breakdown thresholds in laboratory air at standard ambient temperature and pressure, indicates that the ionization energy
is the significant term for fluence thresholds in Eq. \eqref{GrindEQ__2_} rather than work function.

\begin{table}[h]
\caption{ Measured thresholds, $\upvarphi^{\rm exp}_{\rm th}$, ionization energies, $\upvarepsilon_{\rm i}$,  and comparison with tabulated ionization energies, $\upvarepsilon_{\rm tab}$, and work functions, W$_{\rm tab}$, for bulk metallic targets.}
\begin{tabular}{p{0.5in}cccc} \hline\hline
 & $\upvarphi^{\rm exp}_{\rm th}$ & $\upvarepsilon_{\rm i}$ & $\upvarepsilon_{\rm tab}$ & W$_{\rm tab}$ \\
&  (J/cm$^2$)            &  (eV)              &   (eV)                    &   (eV)        \\ \hline
Pb &  0.42 $\pm$ 0.04 \ \ \  &\ \ \ 7.4 $\pm$ 0.2 \ \ \ & 7.4 & 4.1 \\
In &  0.80 $\pm$ 0.05 \ \ \  &\ \ \ 5.8 $\pm$ 0.1 \ \ \ & 5.8 & 4.1 \\
Al &  2.10 $\pm$ 0.07 \ \ \  &\ \ \ 5.9 $\pm$ 0.2 \ \ \ & 5.9 & 4.1 \\
Ag &  2.55 $\pm$ 0.10 \ \ \  &\ \ \ 7.6 $\pm$ 0.3 \ \ \ & 7.6 & 4.7 \\
Cu &  3.40 $\pm$ 0.07 \ \ \  &\ \ \ 7.7 $\pm$ 0.2 \ \ \ & 7.7 & 4.7 \\ \hline\hline
\end{tabular}
\label{table3}
\end{table}
\vspace*{0.125in}

Excellent agreement of measured and tabulated first ionization potentials can be noticed in Table \ref{table3} for the different elements. This agreement can be viewed as a further confirmation of the validity of the laser term as suggested previously in generation of plasma in laboratory air at or near the surface of different target materials \cite{Ref1,Ref2}. This agreement extends to investigations of the laser wavelength dependency of the thresholds, \textit{i.e.}, the predicted and measured threshold fluences agree when using the second harmonic of the Q-switched Nd:YAG laser device at 532 nm.

\section{Conclusion}
The laser induced plasma threshold dependence on laser wavelength and ionization energy was utilized to provide means to measure the first ionization potential of elements. The measured ionization energies for the different targets nicely agree with tabulated values.  In turn, this agreement can be interpreted as confirmation for the laser-induced plasma threshold model.

\section*{Acknowledgment}
The authors acknowledge the continued interest and comments
from Professor Th. M. EL Sherbini and thank for support in part by the
Center for Laser Applications at the University of Tennessee Space Institute
and in part by the Laboratory of Laser and New Materials at Cairo
University.



\begin{thebibliography}{10}
\bibitem{Ref1} A.M. EL Sherbini, C.G. Parigger, ``Wavelength dependency and threshold measurements for nanoparticle-enhanced laser-induced breakdown spectroscopy,'' Spectrochim. Acta B 116, 8--15 (2016).
\bibitem{Ref2} A.M. EL Sherbini, C.G. Parigger, ``Nano-material size dependent laser-plasma thresholds,'' Spectrochim. Acta Part B 124, 79--81 (2016).
\bibitem{Ref3} E.G. Gamaly, A.V. Rode, B. Luther-Davies, ``Ultrafast ablation with high-pulse-rate lasers. Part I: theoretical considerations,'' J. Appl. Phys. 85, 4213-4121 (1999).
\bibitem{Ref4} A.V. Rode, B. Luther-Davies, E.G. Gamaly, ``Ultrafast ablation with high-pulse-rate lasers. Part II: Experiments on laser deposition of amorphous carbon films,'' J. Appl. Phys. 85, 4222--4230 (1999).
\bibitem{Ref5} L.M. Cabalin, J.J. Laserna, ``Experimental determination of laser induced breakdown
thresholds of metals under nanosecond Q-switched laser operation,''  Spectrochim. Acta Part B 53, 723-730 (1998).
\bibitem{Ref6} B.W. Boreham,  J.L. Hughes, ``Measurement of ionization threshold intensities in helium using ponderomotive force accelerated electrons," Sov. Phys. JETP 53, 252--259 (1981).
\bibitem{Ref7} D.X. Hammer, R.J. Thomas, G.D. Noojin, B.A. Rockwell, P.K. Kennedy, W.P. Rooach, ``Experimental Investigation of Ultrashort Pulse Laser-Induced Breakdown Thresholds in Aqueous Media,'' IEEE J Quantum Electron. 32, 670--678 (1996).
\bibitem{Ref8} T.X. Phuoc, Opt. Commun. 175, ``Laser spark ignition: experimental determination of laser-induced breakdown thresholds of combustion gases,'' 419--423 (2000).
\bibitem{Ref9} S. Brieschenk, H. Kleine, S. O'Byrne, ``On the measurement of laser-induced plasma breakdown thresholds,'' J. Appl. Phys. 114, 093101 (2013).
\bibitem{Ref10} A.E. Hussein, P.K. Diwakar, S.S. Harilal, A. Hassanein, ``The role of laser wavelength on plasma generation and expansion of ablation plumes in air,''J. Appl. Phys. 113, 143305 (2013).
\bibitem{Ref11} A. Vogel, K. Nahen, D. Theisen,  J. Noack, ``Plasma Formation in Water by Picosecond and Nanosecond Nd:YAG Laser Pulses—Part I: Optical Breakdown at Threshold and Superthreshold Irradiance,''IEEE J. Sel. Top. Quantum Electron. 2, 847--860 (1996).
\bibitem{Ref12} N. Linz, S. Freidank, X. Liang, H Vogelmann, T. Trickl,  A. Vogel, ``Wavelength dependence of nanosecond infrared laser-induced breakdown in water: Evidence for multiphoton initiation via an intermediate state," Phys. Rev. B 91, 134114 (2015).
\bibitem{Ref13} S. Rothe \textit{et al.}, ``Measurement of the first ionization potential of astatine by laser ionization spectroscopy,'' Nat. Commun. 4, 1835 (2013).
\bibitem{Mulser} P. Mulser, D. Bauer, \textit{High Power Laser-Matter Interaction} (Springer Verlag, Heidelberg, 2010).
\bibitem{Lotz} W. Lotz, ``Ionization Potentials of Atoms and Ions from Hydrogen to Zinc,'' J. Opt. Soc. Am. 57, 873--878 (1967).
\bibitem{Ref14}  A.M. EL Sherbini,  A.M. Aboulfotouh, C.G. Parigger, ``Electron number density measurements using laser-induced breakdown spectroscopy of ionized nitrogen spectral lines,'' Spectrochim. Acta Part B 125, 152--158 (2016).
\bibitem{Ref15} Kramida, A., Ralchenko, Yu., Reader, J. and NIST ASD Team (20l6). NIST Atomic Spectra Database (version 5.4), [Online]. Available: http://physics.nist.gov/asd [Fri Dec 09 2016]. National Institute of Standards and Technology, Gaithersburg, MD; https://www.nist.gov/pml/atomic-spectra-database (last accessed 9/24/2016).
\end{thebibliography}
\end{document}